\documentclass[prl, twocolumn, superscriptaddress]{revtex4-1}
\usepackage{bm, amsmath, amsfonts, amssymb}
\usepackage[italicdiff]{physics}
\usepackage{braket}
\usepackage{comment}
\usepackage{subfigure}
\usepackage{color}
\usepackage{graphicx}

\usepackage{qcircuit}
\usepackage{url}
\usepackage{bm}
\usepackage[breaklinks=true]{hyperref}

\begin{document}
\title{Time-crystalline long-range order in squeezed ground state}

\author{Nobuyuki Okuma}
\email{okuma@hosi.phys.s.u-tokyo.ac.jp}
\affiliation{
 Yukawa Institute for Theoretical Physics, Kyoto University, Kyoto 606-8502, Japan
}

\date{\today}
\begin{abstract}
It is widely believed that ground-state time crystals are not realizable in realistic macroscopic systems.
In particular, Watanabe and Oshikawa proved a theorem that implies the absence of the time-dependent long-range order (TDLRO) in the ground states of short-range many-body systems.
However, this theorem does not forbid the presence of the ground-state TDLRO for macroscopic quantities.
In this work, we investigate a simple bosonic model with a squeezed ground state and point out that the time-dependence of the ground-state TDLRO for the number operator is proportional to the square of the average number in the infinite-squeezing limit, or equivalently, the infinite average-number limit. This result implies the presence of the TDLRO for macroscopic boson number.
We also discuss the physical implementations in optical, spin, and tight-binding systems, including the variants.
We find an example with the macroscopic TDLRO whose essence is low-lying-state physics and another with $marginal$ TDLRO at the quantum critical point.
In addition, we reconsider the definition of the ground-state time crystal in terms of the Floquet picture.

\end{abstract}

\maketitle
Spontaneous symmetry breaking and long-range order have played important roles in various subfields of physics \cite{Tasaki,Altland-Simons}.
Among them, the crystalline order, which is a consequence of the spatial-translation symmetry breaking, has been the most important concept in condensed matter physics.
Since one can also define the translation symmetry in time, a natural question arises: Are the time-translation symmetry breaking (TTSB) or the time-dependent long-range order (TDLRO) allowed?

Recently, Wilczek proposed a time analogy of the crystal, named ``time crystal" \cite{Wilczek}.
However, several papers proved the absence of time crystal defined on ground states of realistic macroscopic systems \cite{Bruno-13,Watanabe-Oshikawa}.
The TTSB for an observable $\Phi$ is obviously impossible because the expectation value of any Heisenberg operator $\Phi(t)=e^{iHt}\Phi e^{-iHt}$, where $H$ is the time-independent Hamiltonian, for the ground state $|\mathrm{GS}\rangle$ does not depend on time \cite{Watanabe-Oshikawa}:
$\langle\mathrm{GS}|\Phi(t)|\mathrm{GS}\rangle=e^{iE_0t}\langle\mathrm{GS}|\Phi|\mathrm{GS}\rangle e^{-iE_0t}=\langle\mathrm{GS}|\Phi|\mathrm{GS}\rangle$, where $E_0$ is the ground-state energy.
Although the impossibility for the TDLRO is a nontrivial question,
Watanabe and Oshikawa proved a theorem that implies the absence of the macroscopic TDLRO in ground states of general short-range interacting systems \cite{Watanabe-Oshikawa}.
More precisely, they showed the following inequality for the time correlation function:
\begin{align}
    \frac{1}{V^2}\left|\langle \mathrm{GS}|\Phi(t)\Phi|\mathrm{GS}\rangle-\langle \mathrm{GS}|\Phi\Phi|\mathrm{GS}\rangle \right|\leq C\frac{t}{V},\label{inequality}
\end{align}
where $V$ is the volume, $\Phi$ is a macroscopic observable operator given by the summation of local operators over the entire volume, and $C$ is a constant that does not depend on $t$ and $V$.
This inequality implies that the time-dependence of the time correlation function for the macroscopic order normalized by the volume goes to zero in the thermodynamic limit for finite time,
Owing to this inequality, the ground-state time crystal in macroscopic sense becomes believed to be impossible. Other possibilities for the TTSB/TDLRO in non-equilibrium matter have been extensively studied in recent years 
\cite{Sacha-15,Khemani,Else,Yao,zhang2017observation,choi2017observation,Sacha_2017,sacha2020time,else2019discrete,khemani2019brief}, while Ref. \cite{guo2021quantum} generalized the definition of the time crystals for the correlation function that is not normalized by the macroscopic number.

In this work, we consider a model with the squeezed ground state and its variants. We point out that the ground-state TDLRO for macroscopic quantity can be constructed in some sense, without contradicting the no-go theorem.

\paragraph{Reinterpretation of TDLRO and TTSB.---}
Before starting the explicit construction of the TDLRO for a macroscopic quantity, we here reconsider the definition of the TDLRO and relate it to the TTSB.
We first revisit trivial and controversial examples with the TDLRO already known.

Actually, one can construct a trivial TDLRO, as pointed out by Watanabe and Oshikawa \cite{Watanabe-Oshikawa}.
As an example, let us consider the following two-level Hamiltonian:
\begin{align}
    H=-\frac{\omega}{2}\sigma_z,\label{trivial}
\end{align}
where $\sigma$ represents the spin Pauli matrix, and $\omega>0$ is the energy gap. 
For this Hamiltonian,  
the time correlation function defined for the $x$-component spin operator $\langle \mathrm{GS}|\sigma_x(t)\sigma_x|\mathrm{GS}\rangle$ exhibits a periodic time dependence $e^{-i\omega t}$ \cite{Watanabe-Oshikawa}. Although this is not the time oscillation in a macroscopic system and ruled out by hand \cite{Watanabe-Oshikawa}, the definition for the TDLRO in Eq. (\ref{inequality}) does not logically forbid it as a zero-dimensional example.
In this sense, it is important to understand this trivial example for seeking the nontrivial TDLRO.
Also, the idea of the trivial TDLRO was generalized to the construction of a controversial example with the macroscopic TDLRO.

Kozin and Kyriienko \cite{Kozin} relaxed the conditions and considered a spin model with long-range infinite-body interactions.
Apart from the details, the essence of this model is nothing but the TDLRO considered in the trivial example.
The only difference is that the states are characterized by the macroscopic quantum number.
The ground state of their model is given by the Schr\"{o}dinger's cat state of total $z$-component spin eigenstates with macroscopic spin quantum number:
\begin{align}
    |\mathrm{GS}\rangle=\frac{1}{\sqrt{2}}\left(|\uparrow\uparrow\cdots\uparrow\rangle+|\downarrow\downarrow\cdots\downarrow\rangle\right).\label{pathological}
\end{align}
Since their models have a finite excitation gap, the TDLRO for the total $z$-component spin oscillates in finite time, as in the case of the Hamiltonian (\ref{trivial}). 
However, this construction is controversial in terms of the pathological infinite-body interactions, as pointed out in Ref. \cite{khemani2020comment}.

Apart from the triviality and the pathological nature, these two examples host the TDLRO at least mathematically. We next elaborate on the reinterpretation of the TDLRO, including these extreme examples.
For convenience, we introduce the Floquet picture usually applied for periodically-driven systems \cite{Oka-19}.
Since the Hamiltonian of interest is static, one can choose an arbitrary period $T$, and the Floquet unitary is nothing but the exponential of the time-independent Hamiltonian:
\begin{align}
    U_{F}:=\exp\left[-i\int^T_0 H(t)\right] =\exp\left[-iTH\right].
\end{align}
For an eigenvalue $u_F$ of the Floquet unitary, the quasi-energy $\epsilon$ is defined as $(i/T)\log u_F$ modulo $2\pi/T$.
In our case, the quasi-energies are nothing but the eigenenergies modulo $2\pi/T$.
In this context, the theory of the ground-state TDLRO with period $0<T<\infty$ for the observable operator $\Phi$, including the trivial and the controversial time crystals, satisfies the followings:
\begin{enumerate}
    \item $|\mathrm{GS}\rangle$ is not an eigenstate of $\Phi$.
    \item The zero quasi-energy eigenstates are degenerated except for the energy-ground-state degeneracy. 
    \item $\Phi$ is closed in the zero quasi-energy eigenspace.
\end{enumerate}
Here we have taken the ground-state energy to be zero.
An eigenstate of $\Phi$ with zero quasi-energy $|\Phi,\epsilon=0\rangle$ is a linear combination of the zero quasi-energy eigenstates, and it has no longer to be an eigenstate of the Hamiltonian or the time-evolution operator $U(t)$, while it is an eigenstate of the Floquet unitary $U_F$. This is nothing but the TTSB state with period $T$. 
Corresponding to the breaking of continuous time translation symmetry, the coset U(1) degrees of freedom indicates huge ``ground-state" degeneracy, as in the case of the ordinary symmetry breaking.
In the case of time, the ``ground-state" degeneracy is defined not for the zero energy but the zero quasi-energy, and the time translation connects different quasi-energy ground states:
$|\Phi^{\rm TTSB}_i\rangle\rightarrow e^{-iHt}|\Phi^{\rm TTSB}_i\rangle$,  without quasi-energy cost.

Using the assumptions, one can express the energy ground state as a superposition of $\Phi$-eigenstates with the TTSB:
\begin{align}
    |\mathrm{GS}\rangle=\sum_{i}A_i|\Phi^{\rm TTSB}_i\rangle,\label{expand}
\end{align}
where $A_i:=\langle\Phi^{\rm TTSB}_i|\mathrm{GS}\rangle$.
This form indicates the time dependence of the correlation function for $\Phi$ with period $T$.
For example, $T=2\pi/\omega$ and $\Phi=\sigma_x$ reproduce the trivial TDLRO. However, such a trivial TDLRO can be constructed for lots of quantum mechanical systems by setting $T=2\pi/(\mathrm{energy~gap})$. 

For the nontrivial TDLRO, we further impose
\begin{align}
     &\langle\mathrm{GS}|\Phi(t)\Phi|\mathrm{GS}\rangle\notag\\
     &=\sum_{i,j}A^*_iA_j\Phi_i\Phi_j\langle \Phi^{\rm TTSB}_i|e^{-iHt}   |\Phi^{\rm TTSB}_j\rangle
     \sim\mathcal{O}(M^2),\label{nontrivialTDlRO}
\end{align}
where $\Phi$ measures the macroscopic quantity $M$ such as a quantity proportional to $V$ in Eq. (\ref{inequality}). For small number of nonzero $A_i$'s, Eq. (\ref{nontrivialTDlRO}) indicates that
the ground state must be a gapped macroscopic cat state of small number of states with $\Phi_i\sim\mathcal{O}(M)$ such as Eq. (\ref{pathological})
in order to observe the macroscopic TDLRO.
In the following, we seek another possibility: 
the squeezed ground state characterized by a superposition of infinite TTSB states.

\paragraph{Squeezed ground state as time crystal.---}
We consider the following bosonic Hamiltonian:
\begin{align}
    H=\omega S(r)a^\dagger aS^\dagger(r),\label{model}
\end{align}
where $\omega>0$ is the frequency, $(a,a^\dagger)$ are the bosonic creation/annihilation operators, and $S(r)$ is the squeeze operator with $r\in \mathbb{R}$:
\begin{align}
    S(r)=\exp\left[\frac{r}{2}(a^2-a^{\dagger2})\right].
\end{align}
The squeeze operator satisfies the following properties \cite{walls2007quantum}:
\begin{align}
    &S^\dagger(r)=S^{-1}(r)=S(-r),\label{unitarity}\\
    &S^\dagger(r) aS(r)=a \cosh r-a^\dagger \sinh r,\label{sas}\\
    &S^\dagger(r) a^\dagger S(r)=a^\dagger\cosh r-a\sinh r.\label{sadags}
\end{align}
Using these properties, one can rewrite the Hamiltonian (\ref{model}) in the quadratic form:
\begin{align}
    H=\omega(a^\dagger\cosh r+a\sinh r)(a\cosh r+a^\dagger\sinh r).\label{quadmodel}
\end{align}
This form implies that the squeezing parameter $r$ represents the degree of non-conserving bosonic number.

The eigenvalue problem of the Hamiltonian (\ref{model}) is easily solved:
\begin{align}
    H[S(r)|N\rangle]=N\omega[S(r)|N\rangle],\label{eigenstate}
\end{align}
where $|N\rangle$ is the eigenstate of the number operator $n=a^\dagger a$ with the eigenvalue $N$. 
Remarkably, the ground state of this model $|\mathrm {GS}\rangle$ is given by the squeezed vacuum state $|0,r\rangle:=S(r)|0\rangle$.

In the following, we calculate the ground-state time correlation function for the number operator \footnote{This quantity was also calculated for a similar model in Ref. \cite{guo2021quantum}, while the perspective of the normalization by macroscopic number was not considered.}:
\begin{align}
    C(t):=\langle \mathrm{GS}|e^{iHt}ne^{-iHt}n|\mathrm{GS}\rangle
    =\langle\mathrm{GS}|ne^{-iHt}n|\mathrm{GS}\rangle.
\end{align}
Using Eqs. (\ref{unitarity},\ref{sas},\ref{sadags},\ref{eigenstate}), we obtain \cite{supplement}
\begin{align}
    C(t)=2(\cosh r)^2(\sinh r)^2e^{-i2\omega t}+(\sinh r)^4.\label{corrfuncsqueeze}
\end{align}
The ground-state time correlation function $C(t)$ has the oscillating term with the frequency $2\omega$, which corresponds to the energy of the bosonic ``Cooper pair" in the squeezed state.

Next, we show that $C(t)$ is macroscopic in terms of the average boson number.
The average of the number operator for the squeezed ground state is calculated as \cite{supplement}
\begin{align}
    N_{\rm ave}&:=\langle \mathrm{GS}|n|\mathrm{GS}\rangle=(\sinh r)^2.\label{naverage}
\end{align}
Thus, we obtain
\begin{align}
    \frac{C(t)}{N_{\rm ave}^2}=2(\coth r)^2e^{-i2\omega t}+1.\label{correp}
\end{align}
In the infinite-squeezing limit ($r\rightarrow \infty$), or equivalently the infinite-number limit ($N_{\rm ave}\rightarrow \infty$), we obtain the following non-vanishing macroscopic TDLRO:
\begin{align}
    \frac{C(t)-C(0)}{N_{\rm ave}^2}\rightarrow2\left[e^{-i2\omega t}-1\right].\label{fraction}
\end{align}
Thus, we insist that the squeezed ground state of the Hamiltonian (\ref{model}) shows the macroscopic time-crystalline order, which is essentially different from the trivial TDLRO in the two-level system.
In our case, the macroscopic quantity is defined as a large but finite number in the semi-infinite-dimensional Fock space for bosons, while 
the definition by Watanabe and Oshikawa uses a quantity proportional to the system volume.
In the spin-implementation part, we compare these two definitions.

\begin{figure}
\begin{center}
 \includegraphics[width=8cm,angle=0,clip]{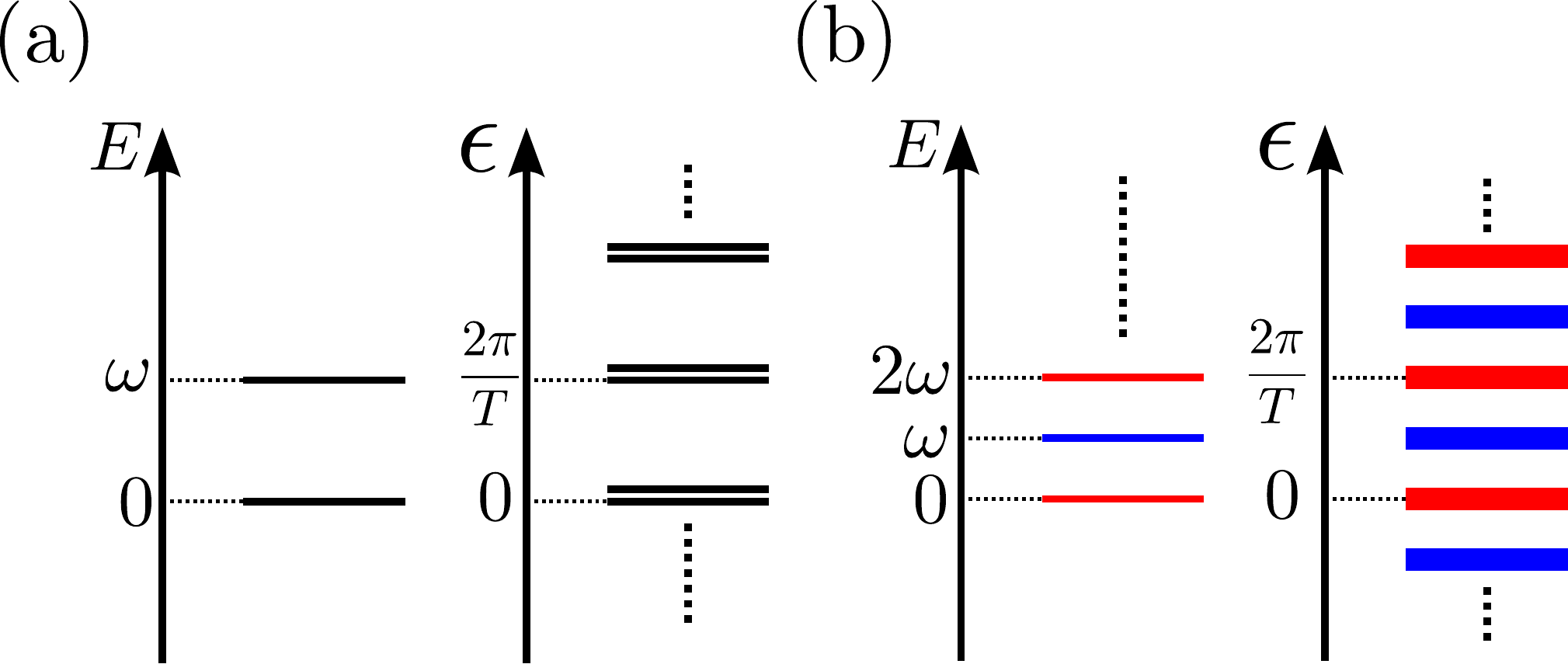}
 \caption{Energy spectra and corresponding quasi-energy spectra.
 In quasi-energy space, the two-level system has two-fold ground-state degeneracy (a), while the zero quasi-energy space for the Hamiltonian (\ref{model}) consists of infinite states with even bosons (b). In general, the energy ground state with the TDLRO can be expressed as a superposition of the quasi-energy TTSB ground states.}
 \label{fig1}
\end{center}
\end{figure}

We note that the ground state considered in this system is a gapped cat state of almost-equally-weighted infinite number-eigenstates with even numbers:
\begin{align}
    |\mathrm{GS}\rangle=\frac{1}{\sqrt{\cosh r}}\sum_n
    \left(\frac{-\tanh r}{2}\right)^n\frac{\sqrt{2n!}}{n!}|2n\rangle.\label{explicityform}
\end{align}
The number operator is closed in the quasi-zero energy space with even boson number (Fig.\ref{fig1}), which ensures the periodicity of the TDLRO.
These number-eigenstates are the TTSB states in Eq. (\ref{expand}) and are charge U(1) invariant, while the TTSB states in Eq. (\ref{pathological}) break the $\mathbb{Z}_2$ symmetry.

We also note that the coherent state, which is also a superposition of infinite number-eigenstates, does not show the macroscopic TDLRO for the number operator.
Let us consider the following bosonic model:
\begin{align}
    H=\omega D(r)a^\dagger aD^\dagger(r),\label{coherentham}
\end{align}
where $D(r)$ is the displacement operator with $r>0$ \cite{walls2007quantum}.
One can solve the eigenvalue problem and calculate the time correlation function in the almost same manner by using the properties for the displacement operator \cite{walls2007quantum}. 
In the infinite-number limit, the macroscopic time oscillation does not exist \cite{supplement}:
\begin{align}
    \frac{C(t)-C(0)}{N_{\rm ave}^2}
    = \frac{e^{-i\omega t}-1}{N_{\rm ave}}\rightarrow0.\label{vanishing}
\end{align}
This difference comes from the degree of the number fluctuation.
While the number uncertainty for the coherent state is determined by the standard quantum limit, those for the squeezed vacuum state can be larger than this limit \cite{walls2007quantum}.

\paragraph{Physical implementation.---}
We here discuss the physical reality of the model (\ref{quadmodel}) and its variants.
Needless to say, photons may be the most promising ingredient for the squeezed ground state because the squeezing technique is widely used in quantum optical techniques \cite{walls2007quantum}, including the effectively-static systems under the rotating-wave approximation.
In the optical implementation, there is no theoretical upper bound for the average number, and thus the macroscopic limit is roughly given by $\coth r\sim1$ (see Eq.(\ref{correp})).

Another possibility is an implementation using the spin squeezed state \cite{Kitagawa-Ueda,ma2011quantum}.
The corresponding spin model is
\begin{align}
    H_{\rm spin}=-h S_z+gS_x^2+\mathrm{Const.},\label{spinham}
\end{align}
where $S_i$'s are macroscopic spin-$S$ operators.
For small spin fluctuation,
by using the Holstein-Primakoff transformation $S_z=S-a^\dagger a,S_{+}=\sqrt{2S}\sqrt{1-a^\dagger a/2S}a\sim\sqrt{2S}a$ \cite{holstein}, where ($a,a^{\dagger}$) are magnon creation/annihilation operators, one can reproduce the Hamiltonian (\ref{quadmodel}) for the parameter set $(h,g)=(\omega e^{-2r},2\omega \sinh r\cosh r/S)$.
In this case, the fluctuation of the
bosonic number is nothing but those of the $z$-component spin.
Once we decompose the macroscopic spin into $N$ $1/2$-spins ($S_i=\sum_x\sigma_i(x)/2$) and allow an arbitrary total spin, the model (\ref{spinham}) can be regarded as an exactly-solvable model with long-range interactions, called Lipkin-Meshkov-Glick model \cite{lipkin1965validity}, which has also been studied in time crystals 
not for the ground states \cite{FloquetLMG,NGStimecrystal}. 
Since the total spin operator commutes with the Hamiltonian, the eigenstates are block-diagonalized in terms of the total spin, and the unique ground state is in a total spin sector $S=N/2$ for $g<0$ (ferromagnetic) and $h>N|g|$, while we can also separate the $S=N/2$ sector from the others by adding the term $\Delta H\propto\sum_i S_i^2$.
In the large-$N$ limit, the squeezing properties are well described by non-interacting Holstein-Primakoff bosons \cite{ma2009fisher}.
In this sense, one can investigate the same physics in the bosonic model defined in the infinite-dimensional Hilbert space.

\begin{figure}[t]
\begin{center}
 \includegraphics[width=8.5cm,angle=0,clip]{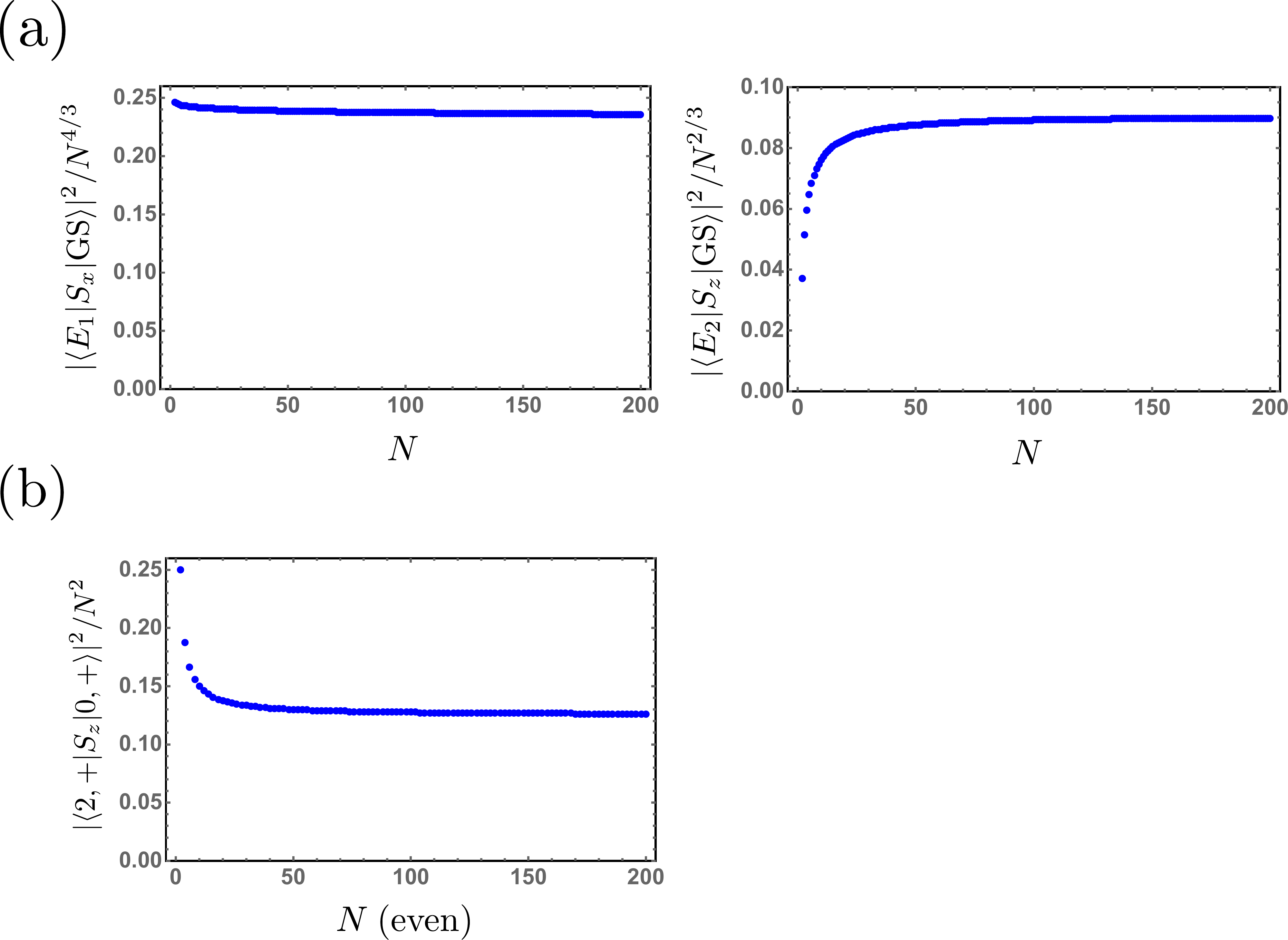}
 \caption{Dominant matrix elements in the time-dependent correlation functions for (a) $h>0,g<0, h=N|g|$ and (b) $h=0,g>0$. $E_1$ and $E_2$ denote the first and second excited states. }
 \label{fig2}
\end{center}
\end{figure}

Next, we discuss the relationship between our TDLRO and the conventional TDLRO in Eq. (\ref{inequality}) for spin squeezed states.
Unlike in the photonic case, there is the upper bound for the spin squeezing, which is achieved when $\max(\langle S_{\perp}^2\rangle) \sim N^2$, where $S_{\perp}$ is a spin component perpendicular to the direction of the spin polarization \cite{Kitagawa-Ueda,ma2011quantum}.
If it is achieved, one can replace $e^{|r|}$ with $\sqrt{N}$, which indicates that our TDLRO is equivalent to the conventional TDLRO.
Unfortunately, this limit cannot be achieved in the Lipkin-Meshkov-Glick model for the ferromagnetic parameters with the unique ground state (symmetry-unbroken phase) \cite{ma2009fisher} owing to the boson-boson interactions. 
At best, $\langle S_{x}^2\rangle\sim N^{4/3}$ is achieved  at the quantum phase transition point $h=N|g|$, which is not described by non-interacting bosons \cite{ma2009fisher}.
In this case, the amplitude of the TDLRO for $S_x$ normalized by $N^2$ is proportional to $1/N^{2/3}$, while that for $S_z$ proportional to $1/N^{4/3}$.
These behaviors are checked by numerically calculating the dominant matrix elements between the ground and excited states with energies proportional to $gN^{2/3}$ \cite{supplement} (Fig. \ref{fig2}). Since the other minor matrix elements are negligibly small \cite{supplement}, the periodicity is also confirmed.
For the constant $g$, the TDLRO for $S_x$ with period $\sim \mathcal{O}(N^{-2/3})$ can break the Watanabe-Oshikawa's inequality (\ref{inequality}) during the time proportional to $N^{1/3}$.
This is not a counterexample for their theory because they assumed the short-range nature, while the present Hamiltonian has long-range interactions.
This ``marginal" example, which has the vanishing amplitude of the  TDLRO but breaks the inequality (\ref{inequality}), is the third possibility for the time crystal, in addition to the trivial and the macroscopic TDLRO. 
We remark that in this example, large fluctuation at the quantum critical point boosts the amplitude of the TDLRO.


To find the macroscopic TDLRO in the conventional sense, we consider another interesting limit: the antiferromagnetic case $(g>0)$ with $g/h\rightarrow \infty$ whose solutions in the $S=N/2$ sector are given for even $N$ by \cite{pan1999analytical}
\begin{align}
    |w,\pm\rangle=\sum_{\rho}&F^{\pm}_N(w,\rho)|\frac{N}{2},\frac{N}{2}-2\rho-0(1)\rangle,\label{exactsol}
\end{align}
where $|S,M_z\rangle$ is an $S_z$-eigenstate with the eigenvalue $M_z$, $|w,+\rangle/|w,-\rangle$ is an energy-eigenstate with $E^{(+)}_{w}/E^{(-)}_{w}=(g/4)w^2/(g/4)(w+2)^2~(w=0,2,\cdots, N)$, and $F^{\pm}_N$ are coefficients \cite{supplement}.
The ground state for even $N$ is unique, and all the excited states have two-fold degeneracy, which is broken by the introduction of finite $h$.
Remarkably, the energy-eigenstates have a similar form as the bosonic squeezed states.
In particular, the ground-state is a cat state with almost-equally-weighted infinite number-eigenstates with the even quantum-number deviation from $S=N/2$,
which mimics Eq.(\ref{explicityform}).
The conventional time correlation function for the $z$-component spin is given by
\begin{align}
    C(t)=\sum_{w\neq0}e^{-iE_w^{(+)}t}|\langle w,+|S_z|0,+\rangle|^2.\label{nonzeroelement}
\end{align}
On the second line, we have used the fact that $S_z$ and $H$ do not mix the $+$ and $-$ sectors, and $\langle 0,+|S_z|0,+\rangle=0$ from the symmetry.
Actually, the matrix element is finite only for $w=2$, which means that the time correlation function has $T=2\pi/g$ periodic oscillation.
By using Eq. (\ref{exactsol}), we obtain the size-dependence of the conventionally-normalized nonzero matrix element (Fig.\ref{fig2}).
This result shows that the conventional ground-state macroscopic TDLRO does exist in the large-$N$ limit.
Again, this behavior in the long-range interacting system does not attack the no-go theorem derived for short-range systems.
This TDLRO can also be interpreted as the position uncertainty of the ground state of a short-range tight-binding model with the same matrix representation \cite{supplement}, which is also a premising implementation.

We note that our construction only needs the two-body (not infinite-body) long-range interactions, which have also been studied in terms of dissipative time crystals \cite{buvca2019non}.
While we have used the $S_z$ basis in order to compare this case with the bosonic squeezing, it is more easily understood in the $S_x$ basis for this limit because the ground state $|0,+\rangle$ and the excited states $|2,\pm\rangle$ are nothing but $|N/2,M_x=0\rangle$ and the combinations of $|N/2,M_x=\pm1\rangle$, respectively.
By performing a textbook calculation, the normalized matrix element is calculated as $N(N+2)/(8N^2)\rightarrow 0.125$, as shown in Fig. \ref{fig2}. Again, one can separate the $S=N/2$ sector from the other sectors to avoid the huge degeneracy by adding two-body terms $\Delta H\propto\sum_i S_i^2$.

\paragraph{Discussion.---}
In both the optical and the spin implementations,
the energy density is increased by the increase of the macroscopic number. In particular, $N$-independent $g$ means that energy per site diverges in the large-$N$ limit, which breaks the extensive property.
Such a non-extensive nature of energy density may be essential to break the inequality.
If we normalize the coefficients of the Hamiltonian by $N$ to satisfy the extensive property, the estimation of the commutation relations in the proof of the no-go theorem \cite{Watanabe-Oshikawa} can be generalized to our cases, which means that the inequality recovers. 
Even in such a case, these examples are still nontrivial because the amplitudes of the TDLRO are not changed, and one can observe the non-vanishing marginal and macroscopic oscillation in time proportional to $N^{1/3}$ and $N$ (not exponential of it) \cite{supplement}.
We also note that these two examples are only stabilized under no perturbation.
In the former case, we have made use of the quantum critical nature of the phase transition point, which indicates that it is only realized for a fine-tuned setup.
In the latter case, the large fluctuation means that the ground state is unhealthy in the context of statistical physics \cite{Tasaki}.
A healthy ``ground state" is defined by a low-lying state \cite{Tasaki} with U(1)-symmetry breaking:  $|\Xi_{+}\rangle=(1+S_z/|\!|S_z|\mathrm{GS}\rangle|\!|)|\mathrm{GS}\rangle/\sqrt{2}=(|0,+\rangle+|2,+\rangle)/\sqrt{2}$, which has a finite expectation value for $S_z$.
This state is not an energy-eigenstate but a TTSB quasi-energy-eigenstate of the Hamiltonian.
Since the macroscopic fluctuation is a necessary condition for the conventional macroscopic TDLRO with periodicity \cite{supplement}, a ground-state macroscopic time crystal should always be ``unhealthy", which requires precise experiments for its realization.

Finally, we remark that if a symmetry-unbroken low-lying state with a long-range order $|\Gamma\rangle=\Phi|\mathrm{GS}\rangle/|\!|\Phi|\mathrm{GS}\rangle|\!|$ \cite{Tasaki} is an excited energy-eigenstate as in the case of $g>0$, then one can observe the ground-state macroscopic TDLRO: $\langle\mathrm{GS}|\Phi e^{-i(H-E_0)t}\Phi|\mathrm{GS}\rangle=\mathcal{O}(V^2)e^{-i\Delta Et}$, where we have used the assumption for the long-range order: $\langle\mathrm{GS}|\Phi^2|\mathrm{GS}\rangle\sim\mathcal{O}(V^2)$. Such low-lying energy-eigenstates have been discussed in the context of Anderson's tower of states \cite{Tasaki}.
For example, it is known that a toy model $H^{\mathrm{toy}}=\bm{S}_A\cdot\bm{S}_B/V$, where $\bm{S}_{A,B}$ are macroscopic spins at sub-lattices $A$ and $B$, hosts tower of eigenstates with eigenenergies $E_m=m(m+1)/V$.
Again, this model is a long-range interacting model, while it describes well low-energy eigenstates of the Heisenberg antiferromagnet \cite{Tasaki}.
Apart from the difference between spin and bosonic models, the second excited eigenstate is given by ``low-lying" state $|\Gamma\rangle$ for the order parameter $\Phi=n-N_{\rm ave}$.


\begin{acknowledgements}
I thank Yuya O. Nakagawa for fruitful discussions.
This work was supported by JST CREST Grant No.~JPMJCR19T2, Japan. N.O. was supported by JSPS KAKENHI Grant No.~JP20K14373.
\end{acknowledgements}

\bibliography{Timecrystal}

\widetext
\pagebreak

\renewcommand{\theequation}{S\arabic{equation}}
\renewcommand{\thefigure}{S\arabic{figure}}
\renewcommand{\thetable}{S\arabic{table}}
\setcounter{equation}{0}
\setcounter{figure}{0}
\setcounter{table}{0}

\begin{center}
{\bf \large Supplemental Material for 
``Time-crystalline long-range order in squeezed ground state"}
\end{center}

\section{SI.~Types of time-dependent long-range order}
\begin{figure}[b]
\begin{center}
 \includegraphics[width=16cm,angle=0,clip]{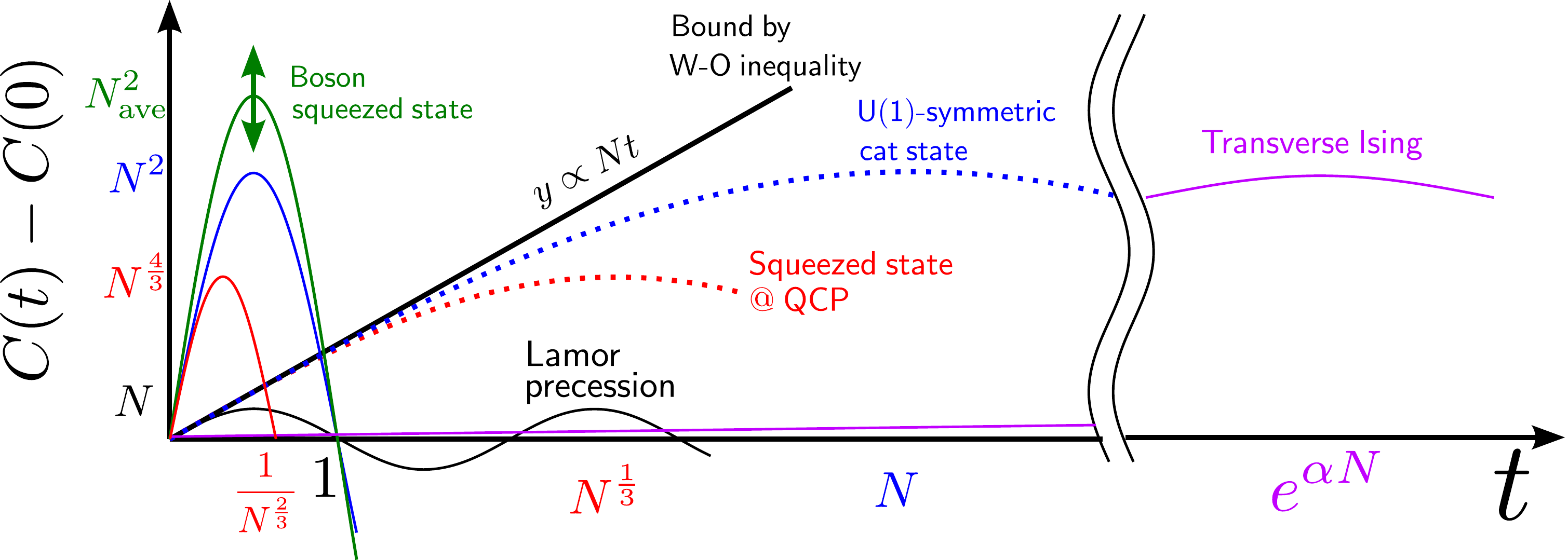}
 \caption{Schematic picture of time-correlation function for various ground states. The dotted lines denote the TDLRO for energy re-scales for the system to be extensive. The Watanabe-Oshikawa's (W-O) inequality is represented as a line in this picture. }
 \label{sup1}
\end{center}
\end{figure}
We here classify the time-dependent long-range order in condenced matter (Fig. \ref{sup1}).
As we discussed in the main text, the two-level system (\ref{trivial}) has a trivial ground-state TDLRO with period $T=2\pi/\omega$.
If we gather $N$ copies of this system, then the amplitude of the ground-state TDLRO for $S_x=\sum^N_i=\sigma_x(i)/2$ is proportional to $N$, and the period is $T=2\pi/\omega$. 
Therefore, we classify such a case as the trivial TDLRO.
In the context of the Hamiltonian (\ref{spinham}) with $g=0$, this TDLRO corresponds to the Lamor precession. 
In terms of the magnon operator, the time-correlation function is also calculated as
\begin{align}
    C(t)=\langle0|\sqrt{\frac{S}{2}}(a+a^{\dagger})e^{-in\omega t}\sqrt{\frac{S}{2}}(a+a^{\dagger})|0\rangle=\frac{N}{4}e^{-i\omega t}.
\end{align}
In the language of bosons, the trivial TDLRO detects the vacuum fluctuation.
The coherent state, including the spin analogy, also has the trivial TDLRO. 

If we are allowed to wait for pathologically long time proportional to exponential of $N$,
the ground state of the transverse Ising model
\begin{align}
    H=-\sum_i\sigma_z (i)\sigma_z (i+1)+h\sum_i\sigma_x (i)
\end{align}
can have the macroscopic TDLRO.
For small but finite $h$, the ground state is a cat state given in Eq. (\ref{pathological}), and the energy gap is exponentially small with respect to $N$.
Owing to this small energy gap, the period of the TDLRO is pathologically long.

In the case of boson squeezing, the amplitude can be arbitrary large, at least theoretically.
Experimentally, on the other hand, the large squeezing is not easy because such a state should contain the large number of photons.
In the cases of the spin squeezed state at the quatnum critical point (QCP) and the U(1)-symmetric cat state in the spin systems, the long-range and non-extensive nature of the Hamiltonian enables them to break the Watanabe-Oshikawa inequality.
Thus, by re-scaling the Hamiltonian by $N$ to be extensive, the inequality holds for such cases.
Even under the re-scaling, one can observe the $marginal$ and the macroscopic TDLRO in time proportional to $N^{1/3}$ and $N$ (not exponential of it).
In both the bosonic and spin cases, an increase of the macroscopic number corresponds to that of the energy density of the physical implementations.

Finally, we remark the relation between the conventional macroscopic TDLRO and the fluctuation for the macroscopic operator $\Phi$:
\begin{align}
    \Delta\Phi^2:=\langle\mathrm{GS}|\Phi\Phi|\mathrm{GS}\rangle-|\langle\mathrm{GS}|\Phi|\mathrm{GS}\rangle|^2=\sum_{m\in\{\mathrm{ES}\}}|\langle m|\Phi|\mathrm{GS}\rangle|^2=C(t=0)-|\langle\mathrm{GS}|\Phi|\mathrm{GS}\rangle|^2.
\end{align}
where \{ES\} is the set of all the excited states.
The macroscopic TDLRO ensures that the sum of some or all matrix elements is proportional to $N^2$.
Thus, $\Delta\Phi^2=\mathcal{O}(N^2)$ is a necessary condition for the macroscopic TDLRO.

\section{SII.~Derivations of several equations}
We here write down the derivations for several quantities.
\subsection{Derivation of Eqs. (\ref{corrfuncsqueeze}) and (\ref{naverage})}
Using Eqs. (\ref{unitarity},\ref{sas},\ref{sadags},\ref{eigenstate}), one can calculate the time correlation function as
\begin{align}
    C(t)=&\langle0|S^\dagger(r)ne^{-iHt}nS(r)|0\rangle
    =\langle0|S^\dagger(r)nS(r)S^\dagger(r)e^{-iHt}S(r)S^\dagger(r)nS(r)|0\rangle\notag\\
    =&\langle0|(a^\dagger\cosh r-a\sinh r)(a \cosh r-a^\dagger \sinh r)e^{-in\omega t}(a^\dagger\cosh r-a\sinh r)(a \cosh r-a^\dagger \sinh r)|0\rangle\notag\\
    =&\left[-\sqrt{2}\cosh r\sinh r\langle2|+(\sinh r)^2\langle0| \right]e^{-in\omega t}\left[-\sqrt{2}\cosh r\sinh r|2\rangle+(\sinh r)^2|0\rangle \right]\notag\\
    =&2(\cosh r)^2(\sinh r)^2e^{-i2\omega t}+(\sinh r)^4.
\end{align}
The average number is simply calculated as
\begin{align}
    N_{\rm ave}:=\langle \mathrm{GS}|n|\mathrm{GS}\rangle=\langle0|S^\dagger(r)nS(r)|0\rangle=\langle0|(a^\dagger\cosh r-a\sinh r)(a \cosh r-a^\dagger \sinh r)  |0\rangle=(\sinh r)^2.
\end{align}

\subsection{Derivation of Eq. (\ref{vanishing})}
The displacement operator $D(r)$ satisfies \cite{walls2007quantum}
\begin{align}
    &D^\dagger(r)=D^{-1}(r)=D(-r),\\
    &D^\dagger(r) aD(r)=a+r,\\
    &D^\dagger(r) a^\dagger D(r)=a^\dagger+r.
\end{align}
The explicit form of the Hamiltonian (\ref{coherentham}) is given by
\begin{align}
    H=\omega D(r)a^\dagger D^\dagger(r)D(r) aD^\dagger(r)
    =\omega(a{^\dagger}-r)(a-r).
\end{align}
Using these equations, we obtain
\begin{align}
    C(t):=&\langle \mathrm{GS}|e^{iHt}ne^{-iHt}n|\mathrm{GS}\rangle
    =\langle\mathrm{GS}|ne^{-iHt}n|\mathrm{GS}\rangle=\langle0|D^\dagger(r)nD(r)D^\dagger(r)e^{-iHt}D(r)S^\dagger(r)nD(r)|0\rangle\notag\\
    &=\langle0|(a^{\dagger}+r)(a+r)e^{-in\omega t}(a^{\dagger}+r)(a+r)|0\rangle=r^2e^{-i\omega t}+r^4.
\end{align}
The average number is simply calculated as
\begin{align}
    N_{\rm ave}:=\langle \mathrm{GS}|n|\mathrm{GS}\rangle=\langle0|D^\dagger(r)nD(r)|0\rangle=\langle0|(a^{\dagger}+r)(a+r)  |0\rangle=r^2.
\end{align}
\section{SIII.~Coefficients in Eq. (\ref{exactsol})}
The complete form of Eq. (\ref{exactsol}) is given by \cite{pan1999analytical}
\begin{align}
    |w,+\rangle\propto\sum_{\mu,\rho}&\frac{[(N-2\rho)!(2\rho)!]^{1/2}(-1)^{\rho-\mu}}{(w-2\mu)!(2\mu)!\left(\frac{N-w}{2}-\rho+\mu\right)!(\rho-\mu)!}|\frac{N}{2},\frac{N}{2}-2\rho\rangle,\notag\\
    |w,-\rangle\propto\sum_{\mu,\rho}&\frac{[(N-2\rho-1)!(2\rho+1)!]^{1/2}(-1)^{\rho-\mu}}{(w+1-2\mu)!(2\mu+1)!\left(\frac{N-w-2}{2}-\rho+\mu\right)!(\rho-\mu)!}|\frac{N}{2},\frac{N}{2}-1-2\rho\rangle.
\end{align}

\section{SIV.~Numerical check of size-dependent behaviors}
\begin{figure}[t]
\begin{center}
 \includegraphics[width=16cm,angle=0,clip]{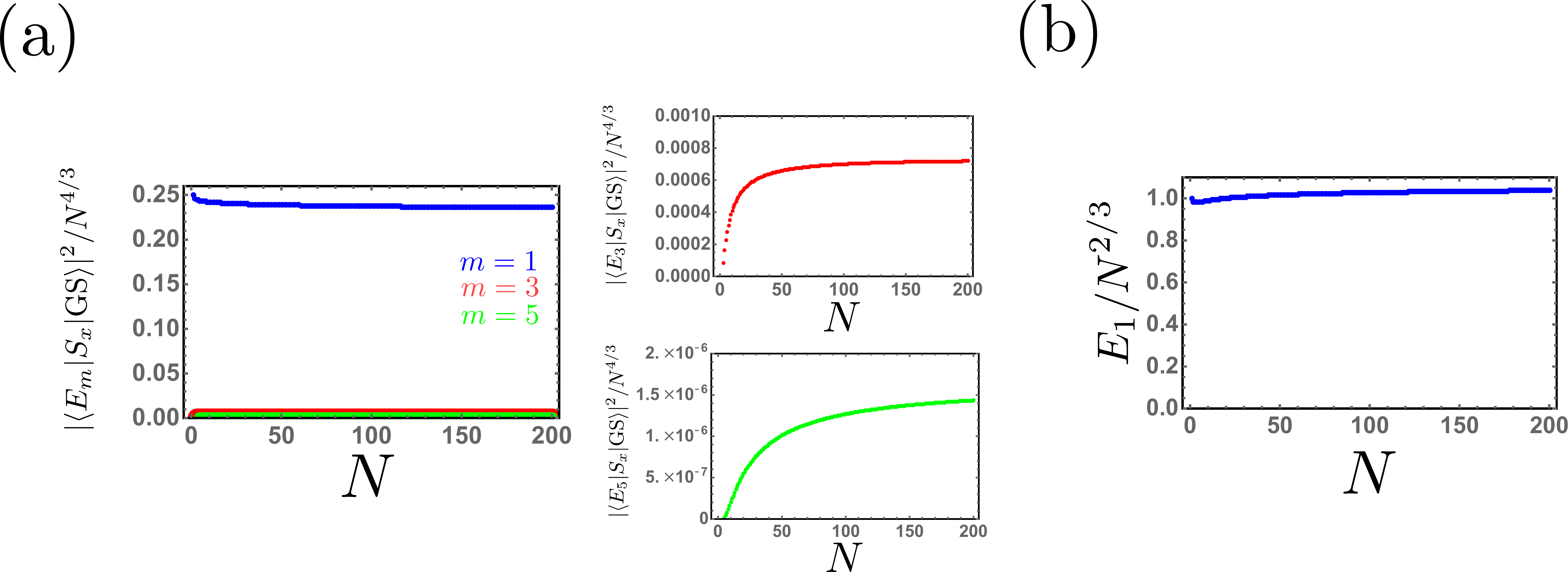}
 \caption{Size-dependent quantities defined for the model (\ref{spinham}) at the critical point ($h=N$ and $g=-1$). (a) Size-dependence of nonzero dominant matrix elements in time-correlation functions. The matrix element between the ground and the first excited states is dominant, and the others are negligibly small, which ensures the periodicity of the $marginal$ TDLRO.  (b) Size-dependence of energy gap.}
 \label{sup2}
\end{center}
\end{figure}

We here discuss the size-dependent quantities defined for the model (\ref{spinham}) at the critical point ($h=N$ and $g=-1$):
\begin{align}
    H_{\rm spin}=-h S_z+gS_x^2+\mathrm{Const.},
\end{align}
where the constant part is set such that the ground-state energy is zero.
The time-dependent correlation function for $S_x$ is expanded in terms of the energy-eigenstates:
\begin{align}
    C(t):=\langle \mathrm{GS}|S_xe^{-iHt}S_x|\mathrm{GS}\rangle
    =\sum_{m}e^{-iE_mt}|\langle E_m|S_x|\mathrm{GS}\rangle|^2,
\end{align}
where $E_0<E_1\cdots$ denote the ground and excited states.
Unlike in the region that is described by non-interacting magnons, $E_m/E_1$ for $m>1$ are no longer integer.
Thus, the ratio between the amplitude of the dominant term and that of the other terms is an important issue to judge whether the correlation function can be regarded as periodic function of time. The size-dependence of the dominant three terms are plotted in Fig. \ref{sup2} (a).
All of them are proportional to $N^{4/3}$ in the large-$N$ limit.
Apparently, the second and the third dominant terms are negligibly small with respect to the most dominant term.

The energy gap between the ground and the first excited state does also depend on the system size.
As shown in Fig. \ref{sup2} (b), the energy gap is proportional to $N^{2/3}$.

\section{SV.~Tight-binding picture of the time-crystalline order in squeezed ground state}
In the main text, we have made use of the large fluctuation from the number-phase uncertainty relation of the squeezed state.
We here discuss the same physics for the short-range tight-binding Hamiltonian that has the same matrix representation as $H=S_x^2$ in the $S_z$ basis:
\begin{align}
    &H=\sum_{i,j}H_{i,j}c_i^{\dagger}c_j,\notag\\
    &H_{i,i+2}=H_{i+2,i}=\frac{1}{4}\sqrt{\left(\frac{N}{2}-i\right)\left(\frac{N}{2}+i+1\right)}\sqrt{\left(\frac{N}{2}-i-1\right)\left(\frac{N}{2}+i+2\right)},\notag\\
    &H_{i,i}=\frac{1}{4}\left(\frac{N}{2}-i\right)\left(\frac{N}{2}+i+1\right)+\frac{1}{4}\left(\frac{N}{2}+i\right)\left(\frac{N}{2}-i+1\right),\label{tbmodel}
\end{align}
where the even number $N$ is the system size (minus one), and $(c_i,c_i^{\dagger})$ are creation/annihilation operators for a particle at site $-N/2\leq i\leq N/2$.
Suppose that there is one particle in the whole system.
Then, the energy spectrum of the system is given by the eigenvalues of the matrix $H_{i,j}$.
In this language, the $S_z$ operator is mapped to the position operator:
\begin{align}
    x=\sum_{i}ic_i^{\dagger}c_i.\label{position}
\end{align}
The time correlation function for the position operator is given by 
\begin{align}
    C(t)/(N+1)^2=\langle \mathrm{GS}|xe^{-i(H-E_0)t}x|\mathrm{GS}\rangle/(N+1)^2=e^{-it}N(N+2)/8(N+1)^2\rightarrow0.125e^{-it}.
\end{align}
Thus, the position of the fermion shows the macroscopic TDLRO.
Since the Hamiltonian and the position operators consist of local operators, it seems that this example breaks the Watanabe-Oshikawa's inequality.
In this case, however, such an expectation is not true.
In general, the constant $C$ in the inequality (\ref{inequality}) depends on the Hamiltonian and the order operator \cite{Watanabe-Oshikawa}. Since the present Hamiltonian depend on the system size $N$, $C$ can depend on $N$. This is the origin of the macroscopic TDLRO in this tight-binding model.

We remark that a similar ``breakdown" of the inequality can occur for the case where the Hamiltonian or/and the order operator consist of the position-dependent local operators.
In such cases, the Hamiltonian/order operator can implicitly depend on the system size, as in Eq. (\ref{position}).
If we parametrize the model (\ref{tbmodel}) by two-dimensional coordinate $\bm{x}=(x,y)$ with $x+y=N+1$, we obtain the position-dependent Hamiltonian:
\begin{align}
    H=\sum_{\{\bm{x}|x+y=N+1\}}\frac{\sqrt{xy(x-1)(y+1)}}{4}(c_{\bm{x}}^{\dagger}c_{\bm{x}+\bm{\delta}}+c_{\bm{x}+\bm{\delta}}^{\dagger}c_{\bm{x}})+\frac{xy+(x+1)(y-1)}{4}c_{\bm{x}}^{\dagger}c_{\bm{x}},
\end{align}
where $\bm{\delta}=(2,-2)$.
Now the Hamiltonian lives on a one-dimensional subspace of the two-dimensional space.
In this parametrization, the coefficients no longer depend on the system size explicitly. 
Apparently, this model also has the macroscopic TDLRO for $x$-component of the position operator.
\end{document}